\documentclass{desyproc}
\usepackage{xspace}
\newcommand{\psq    }{\ensuremath{P^{2}}\xspace}
\newcommand{\qsq    }{\ensuremath{Q^{2}}\xspace}
\newcommand{\ft     }{\ensuremath{F_{2}^{\gamma}}\xspace}
\newcommand{\ftxq   }{\ensuremath{\ft(x,\qsq)}\xspace}
\newcommand{\ftc    }{\ensuremath{F_{2,\mathrm{c}}^{\gamma}}\xspace}
\newcommand{\fl     }{\ensuremath{F_\mathrm{L}^{\gamma}}\xspace}
\newcommand{\flxq   }{\ensuremath{\fl(x,\qsq)}\xspace}
\newcommand{\ftqed  }{\ensuremath{F_\mathrm{2,QED}^{\gamma}}\xspace}
\newcommand{\ftq    }{\ensuremath{\ft(\qsq)}\xspace}
\newcommand{\gevsq  }{\ensuremath{\mathrm{GeV^2}}\xspace}
\newcommand{\msb    }{\ensuremath{\scriptstyle\overline{\rm MS}}\xspace}
\newcommand{\lamv   }{\ensuremath{\Lambda_{\rm 4}^{\msb}}\xspace}
\begin{document}
%
%
\title{Experimental Review of Photon Structure Function Data}
\author{{\slshape Richard Nisius}\\[1ex]
         Max-Planck-Institut f\"ur Physik (Werner-Heisenberg-Institut), 
         F\"ohringer Ring 6, D-80805 M\"un\-chen, Germany, 
         E-mail: Richard.Nisius@mpp.mpg.de\thanks{Invited talk 
         presented at the Photon09 Conference in Hamburg on
         May 12, 2009.}%
        }

\contribID{23}
%
\confID{1407}  
\desyproc{DESY-PROC-2009-03}
\acronym{PHOTON09} 
\doi  
\maketitle
\vspace{-6cm}\begin{flushright}{\bf MPP-2009-131}\end{flushright}\vspace{5cm}
%
%
\begin{abstract}
 The present knowledge of the structure of the photon is presented
 based on results obtained by measurements of photon structure
 functions at e$^+$e$^-$ collider. 
 Results are presented both for the QED structure of the photon as
 well as for the hadronic structure, where the data are also compared
 to recent parametrisations of the hadronic structure function \ftxq.
 Prospects of future photon structure function measurements,
 especially at an International Linear Collider are outlined.
\end{abstract}
%
%
\section{Introduction}
\label{sec:intro}
 The measurements of photon structure functions have a long tradition
 since the first of such measurements was performed by the PLUTO
 Collaboration in 1981.
 The investigations concern the QED structure of the photon as well as
 the hadronic structure. 
 For the hadronic structure function \ftxq the main areas of interest
 are the behavior at low values of $x$ and the evolution with the
 momentum scale \qsq, which is predicted by QCD to be logarithmic.
 The experimental information is dominated by the results from the
 four LEP experiments.
 \par
 This review is based on earlier work~\cite{NIS-9904,NIS-0201} and as
 an extension provides a number of updated figures, together with a
 comparison of the experimental data with new parametrisations of
 \ftxq that became available since then.
 Only results on the structure of quasi-real photons are discussed
 here. The structure of virtual photons and the corresponding
 measurements of effective structure functions are detailed
 in~\cite{SAS-0901}.
%
%
\section{Structure function measurements}
\label{sec:struc}
 The photon can fluctuate into a fermion--anti-fermion state
 consistent with the quantum numbers of the photon and within the
 limitations set by the Heisenberg uncertainty principle. 
 These fluctuations are favored, i.e.~have the longest lifetimes, for
 high energetic photons of low virtuality. If such a fluctuation of
 the photon is probed, the photon reveals its structure.
%
\begin{figure}[thb]
 \centerline{\includegraphics[width=0.50\textwidth]{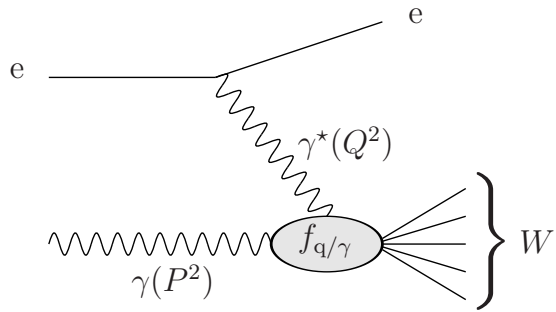}}
 \caption{A sketch of the deep-inelastic electron-photon scattering process.}
 \label{Fig:fig01}
\end{figure}
%
 Using this feature, measurements of photon structure functions are
 obtained from the differential cross-section of the deep-inelastic
 electron-photon scattering\footnote{In this paper, the term electron
 encompasses positrons throughout.} process sketched in
 Figure~\ref{Fig:fig01}.
 In this process the structure of the quasi-real photon, $\gamma$,
 radiated off an electron from one beam is probed by the virtual
 photon, $\gamma^\star$. The $\gamma^\star$ is radiated off an
 electron from the other beam such that this electron is deflected
 into the detector.

 The detailed formalism for the scattering of photons of arbitrary
 virtualities can be found in~\cite{NIS-9904}.
 For deep-inelastic electron-photon scattering on quasi-real photons
 the equation reduces to the well known formula:
%
\begin{equation*}
 \frac{d^2\sigma_{{\rm e}\gamma\rightarrow {\rm e} X}}{dxdQ^2}
 = \frac{2\pi\alpha^2}{x\,Q^{4}}
     \left[\left( 1+(1-y)^2\right) \ftxq - y^{2} \flxq\right]
 \mbox{     with:    }
 x = \frac{Q^{2}}{P^{2}+Q^{2}+W^{2}}
\end{equation*}
%
 The absolute values of the four momentum squared of the virtual and
 quasi-real photons are denoted \qsq and \psq, with $\psq \ll\qsq$.
 The symbols $x$ and $y$ denote the usual dimensionless variables of
 deep-inelastic scattering, $W$ denotes the invariant mass of the
 final state excluding the electrons, and $\alpha$ is the fine
 structure constant.
 The flux of the incoming photons, $f_{\gamma}(z,\psq)$, where $z$ is
 the fraction of the electron energy carried by the photon, is usually
 taken from the equivalent photon approximation, EPA.
 At leading order, the structure function \ftxq is proportional to the
 parton content, $f_{q/\gamma}$, of the photon, and therefore reveals
 the structure of the photon.
 In the region of small $y$ studied, $y \ll 1$, the contribution of
 the term containing \flxq is small, and is usually neglected.
%
%
\subsection{QED structure}
\label{sec:qed}
\begin{figure}[thb]
 \centerline{\includegraphics[width=0.50\textwidth]{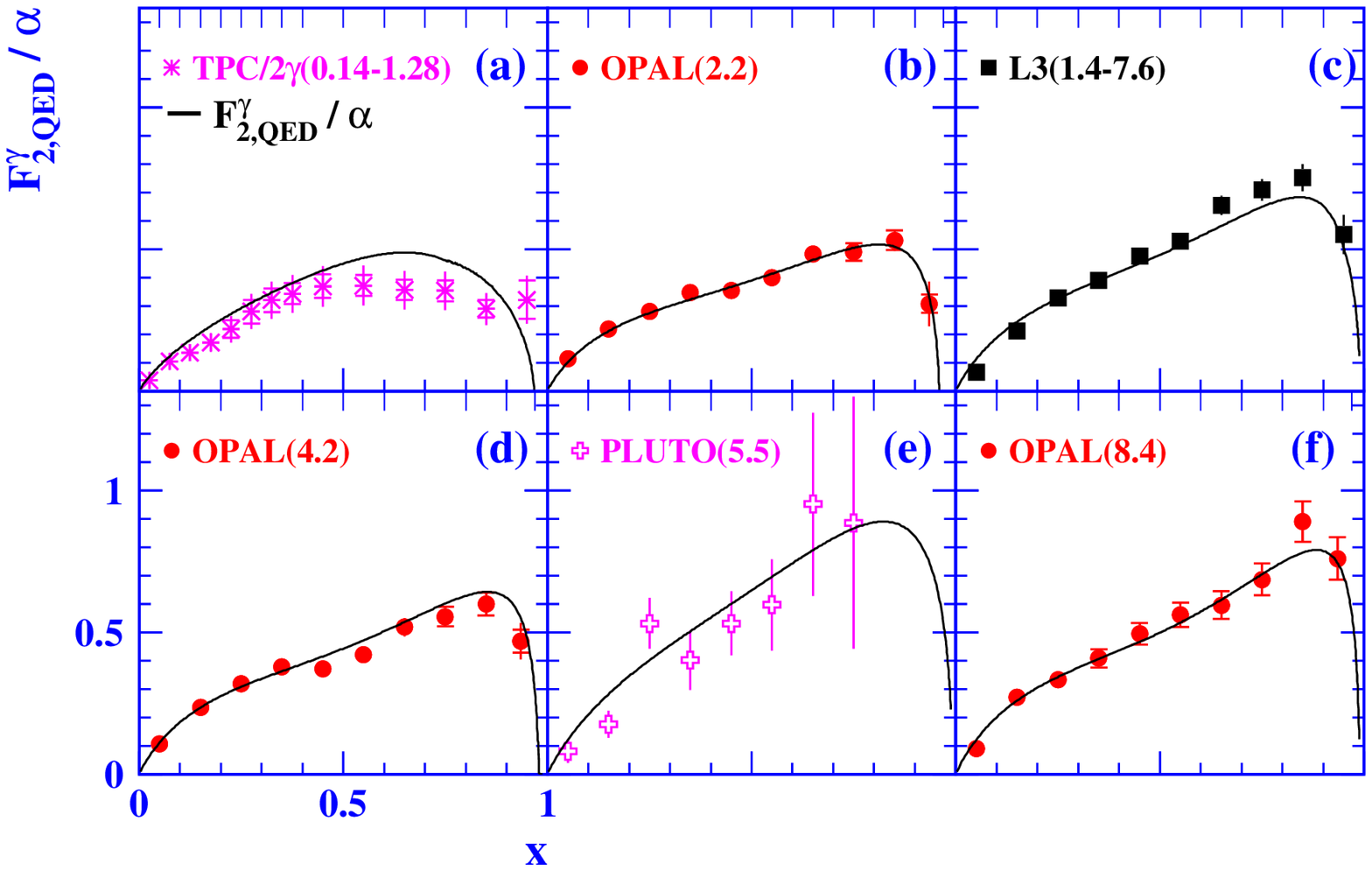}
             \includegraphics[width=0.50\textwidth]{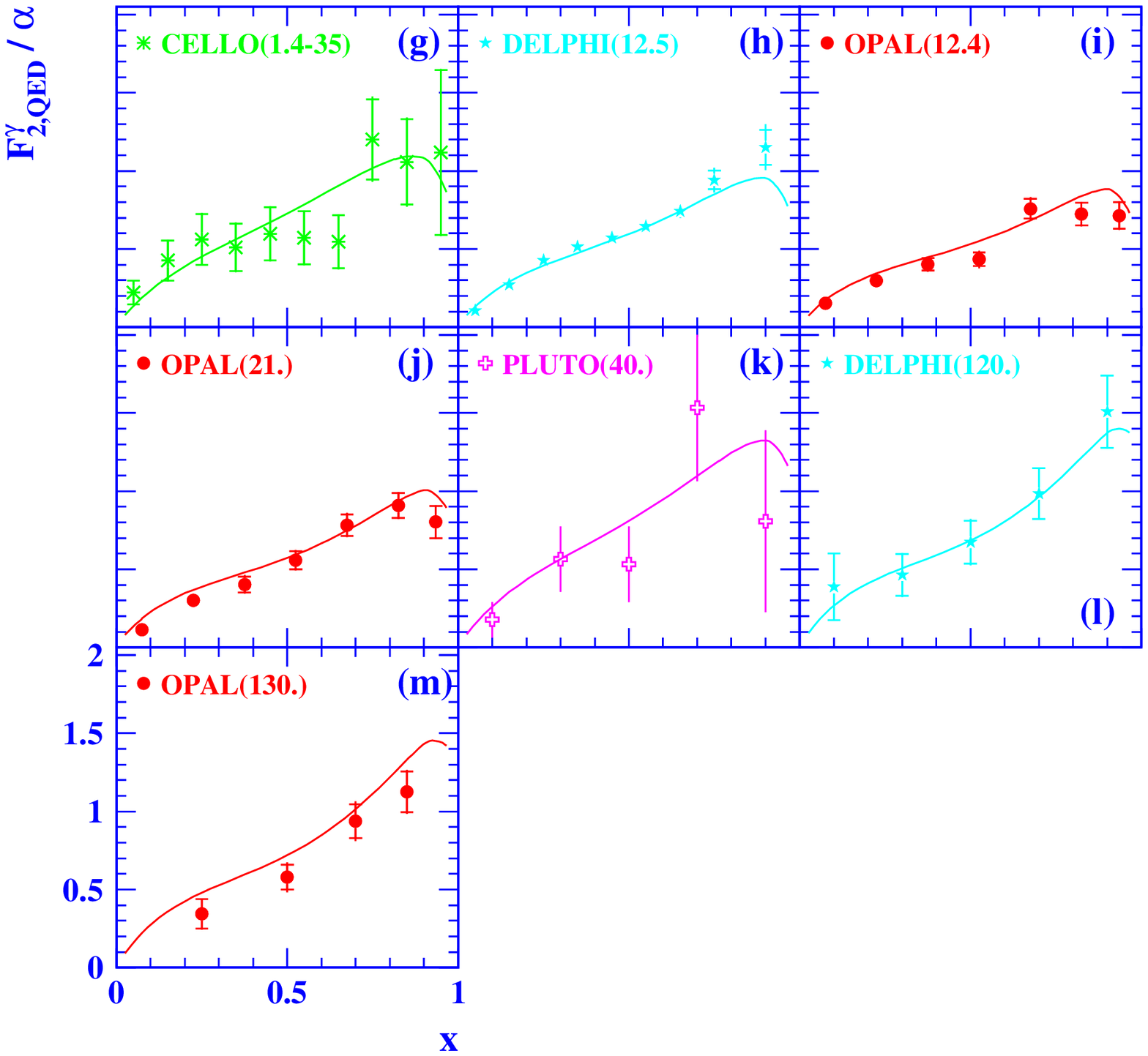}}
 \caption{The world date on the QED structure function \ftqed of the photon.}
 \label{Fig:fig02}
\end{figure}

 The QED structure function \ftqed of the photon is measured from
 deep-inelastic electron-photon scattering events in which a pair of
 muons is produced by the $\gamma\gamma^\star$ system.
 Figure~\ref{Fig:fig02} shows the present world data on this
 measurement. An update is expected when the ongoing L3
 analysis~\cite{DEH-0901} is finalized.
%
%
 The data span a range of about two orders of magnitude in \qsq and
 have a precision down to about 5$\%$.
 With this precision, the treatment of the small but non-zero
 virtuality of the quasi-real photon is important, as are electroweak
 radiative corrections to the deep inelastically scattered electron.
 Unfortunately, the treatment of these corrections is different for
 the various experiments, see~\cite{NIS-9904} for details.

 In addition to the measurements of \ftqed further structure
 functions~\cite{SEY-9801} have been obtained by analyzing the
 azimuthal correlation between the scattering plane of the deep
 inelastically scattered electron and the plane spanned by the muon
 pair. Good agreement between data and predictions has been found.
 Also the scattering of two highly virtual photons has been analyzed,
 and an indirect evidence for the presence of interference terms has
 been found~\cite{OPALPR271}.
 Both these measurements are discussed in detail in~\cite{NIS-9904}.
 \par
 Apart from shining light on the QED structure of the photon the
 determination of the QED structure functions of the photon serves two
 experimental purposes.
 It is a test bed for preparing the tools for the measurements of
 \ftxq, and it sets the limit of precision that could possibly be
 obtained in the more complex case of hadronic final states.
%
%
\subsection{Hadronic structure}
\label{sec:had}
 The measurement of the hadronic structure of the photon is hampered
 by the fact that for measuring $x$, the invariant mass W of the
 hadronic final state has to be reconstructed.
 This is because the energy of the incoming quasi-real photon is not
 known, and consequently, reconstruction of $x$ from the
 deep-inelastically scattered electron alone is impossible.
 Since the hadronic state is not perfectly described by the available
 Monte Carlo models, and part of the final state hadrons are scattered
 into the forward regions of the detectors which are only equipped
 with electromagnetic calorimeters, or even outside of the detector
 acceptance, the precision with which $x$ can be obtained is limited,
 especially at large values of $W$ and correspondingly low values of
 $x$.
 At large values of \qsq the value of $x$ is determined much more
 precisely, and also the data are much better described by the Monte
 Carlo models.
 The problems at low values of $x$ are partly overcome by
 sophisticated unfolding techniques, and by constraining the Monte
 Carlo Models by utilizing combined LEP data on the hadronic final
 state~\cite{OPALPR316}.
 Still the Monte Carlo description at low values of $x$ is one of the
 dominant uncertainties in this measurement, such that some LEP
 experiments even refrained from assigning an error due to this model
 dependence, but published results for individual models instead.
 \par
%
\begin{figure}[t]
 \centerline{\includegraphics[width=0.85\textwidth]{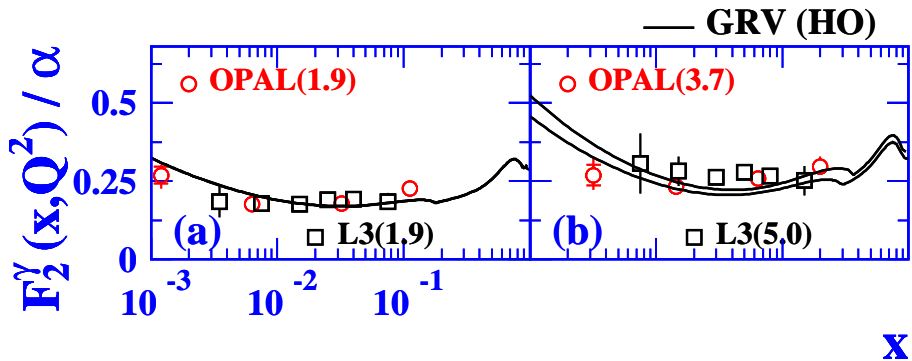}}
 \caption{The world data on \ftxq unfolded on a logarithmic $x$ scale.}
 \label{Fig:fig03}
\end{figure}
%
\begin{figure}[thb]
 \centerline{\includegraphics[width=\textwidth]{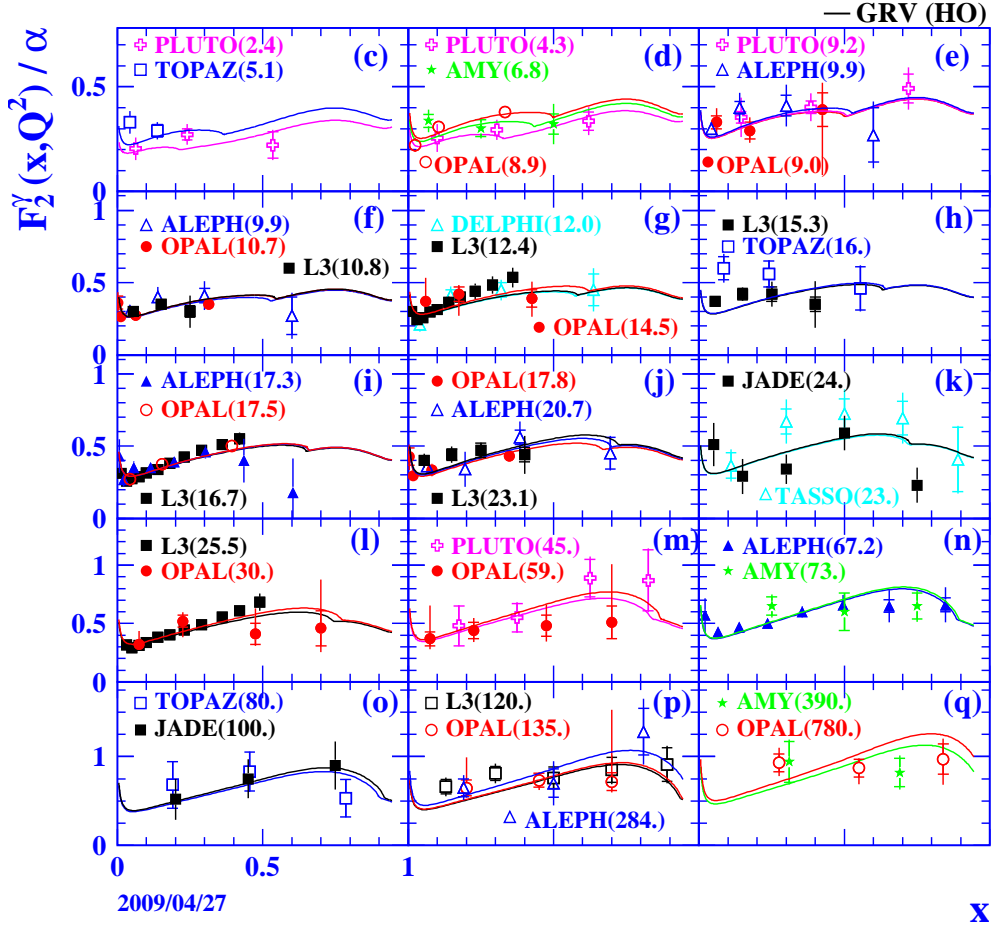}}
 \caption{The world data on \ftxq unfolded on a linear $x$ scale.}
 \label{Fig:fig04}
\end{figure}
%
 There is one important difference between the structure function
 $F_2^p$ of the proton and \ftxq of the photon, which originates in
 the different evolution equations the two have to obey.
 Whereas $F_2^p$ results from a solution of a homogeneous evolution
 equation, the photon structure function \ftxq follows an
 inhomogeneous evolution equation, and therefore receives two
 contributions.
 These are frequently called the hadron-like component, stemming from
 the general solution of the homogeneous evolution equation as for
 $F_2^p$, and the point-like component, resulting from a specific
 solution of the inhomogeneous evolution equation.
 This results in different scaling violations of $F_2^p$ and \ftxq.
 \par
 The present status of the measurements of \ftxq is shown in
 Figures~\ref{Fig:fig03} and~\ref{Fig:fig04}.
 Starting from the data used in~\cite{NIS-9904} the TPC/2$\gamma$
 results are dropped. This is due to their unusual shape as a function
 of $x$ for low values of \qsq, and consequently very bad
 $\chi^2/\mathrm{dof}$ values wrt.~several parametrisations of \ftxq,
 see Tables 4 and 5 in~\cite{NIS-9904}.
 In addition, all preliminary and not yet published LEP data have been
 excluded, and the newly published data from ALEPH~\cite{ALE-0301} and
 L3~\cite{L3C-0001,L3C-0501} have been added.
 The data span a region in \qsq from 1.9--780~\gevsq and in $x$ from
 0.0025--0.98. 
 The experimental precision is clearly dominated by the results from
 the LEP experiments. There is a nice consistency between the results
 obtained at LEP1 energies (open symbols) with the ones from LEP2
 energy data (filled symbols), which at the same \qsq illuminate
 different detector parts.
 The higher order parametrisation from the GRV group~\cite{GLU-9202},
 which has been obtained before many of the shown datasets became
 available, still gives a fair description of the data.
 \par
 Since the end of LEP there has been quite some effort made in
 obtaining new parametrisations of \ftxq by several groups of authors,
 namely CJK~\cite{COR-0401}, AFG~\cite{AUR-0501} and
 SAL~\cite{SLO-0601}.
 Some important ingredients of the various theoretical analyses are
 given below, for further details the reader is referred to the
 original publications.
 \par
%
\begin{figure}[thb]
 \centerline{\includegraphics[width=0.50\textwidth]{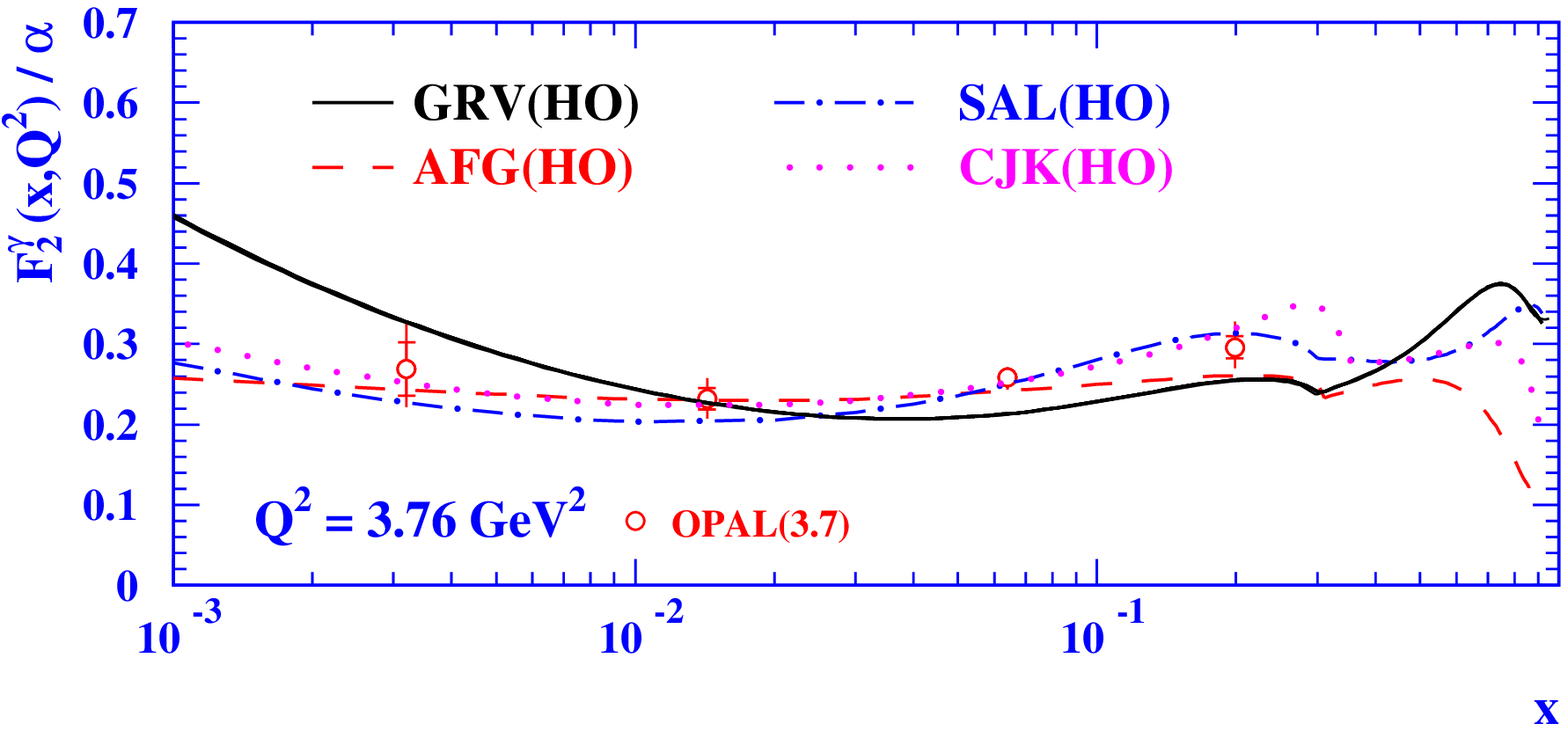}
             \includegraphics[width=0.50\textwidth]{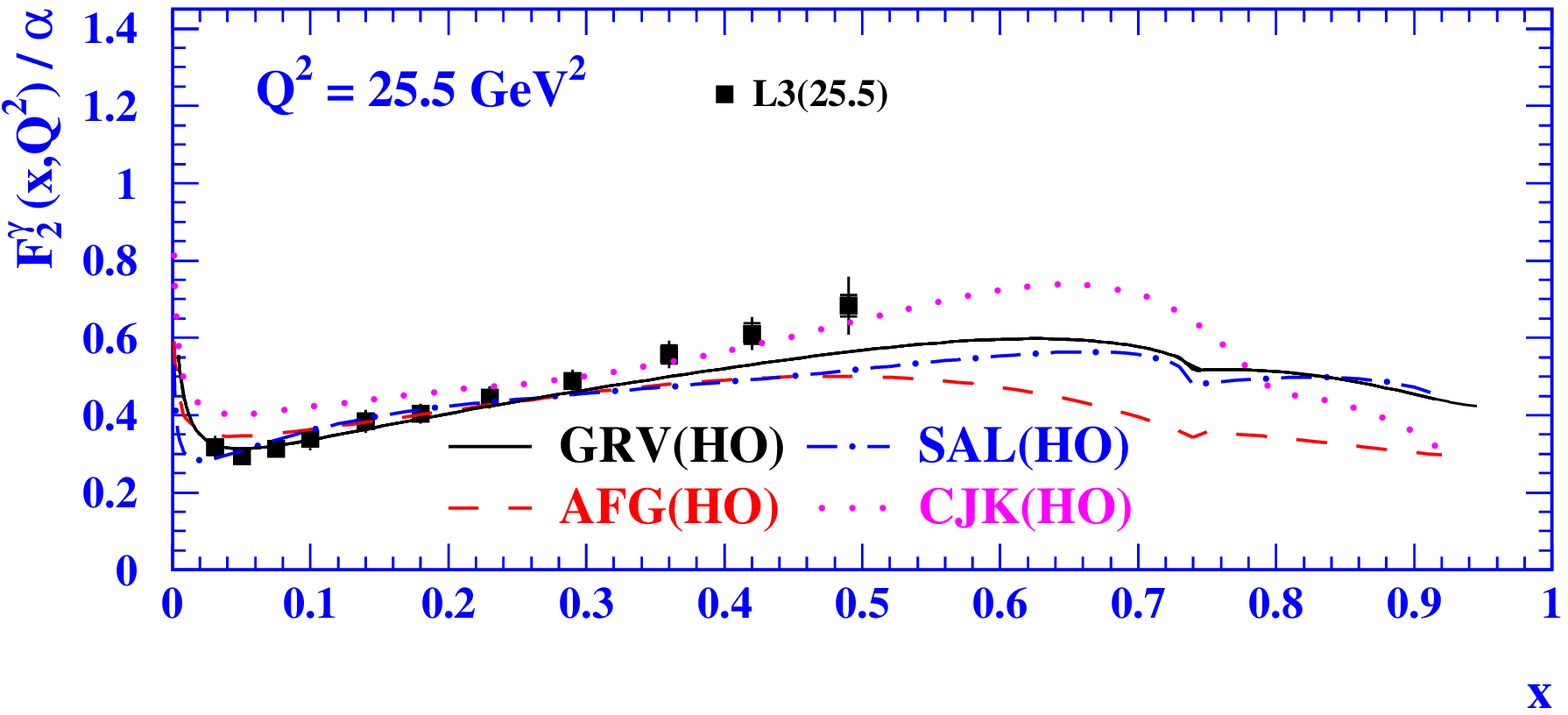}}
 \caption{Parametrisations of \ftxq compared to LEP data at 
          $\qsq=3.7$ and $25.5$ \gevsq.}
 \label{Fig:fig05}
\end{figure}
%
%
\begin{figure}[thb]
 \centerline{\includegraphics[width=0.50\textwidth]{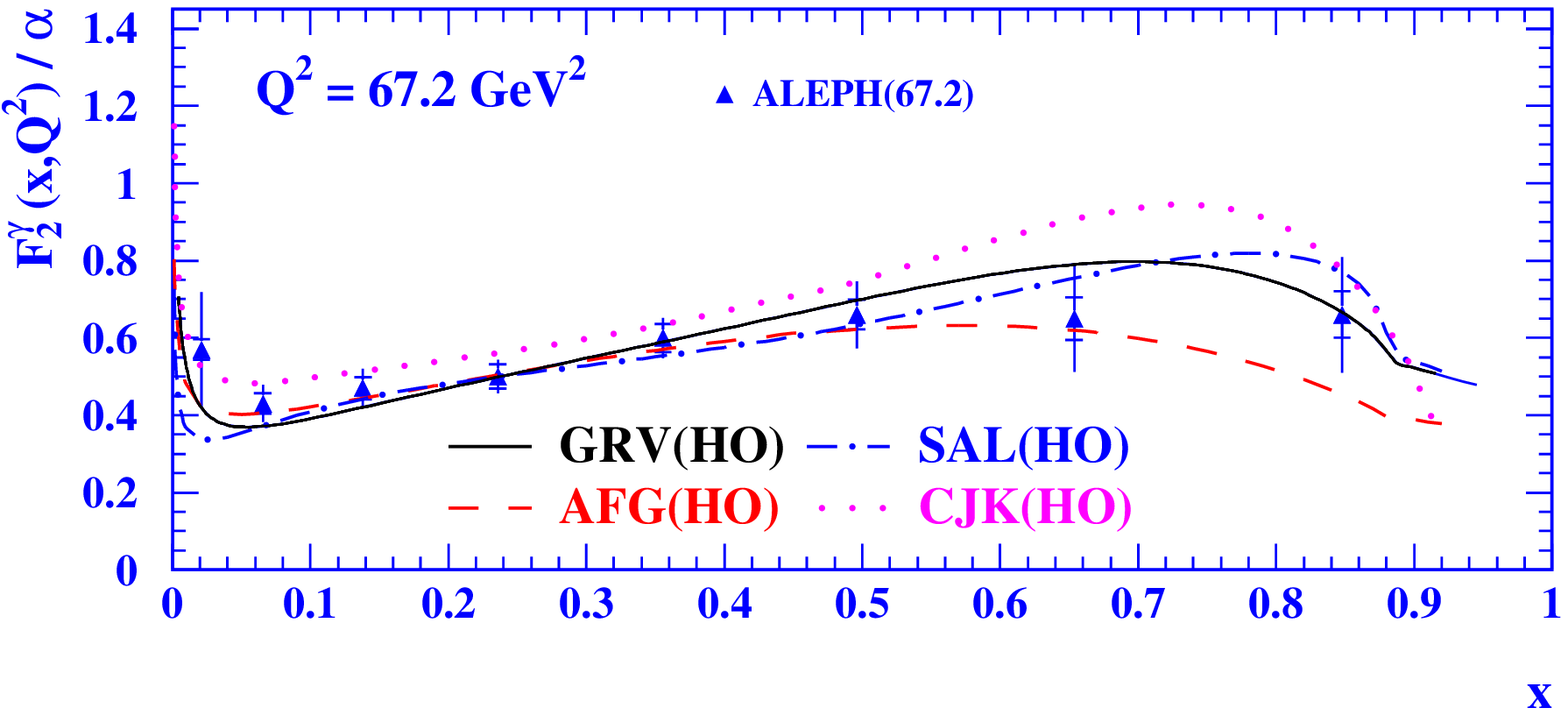}
             \includegraphics[width=0.50\textwidth]{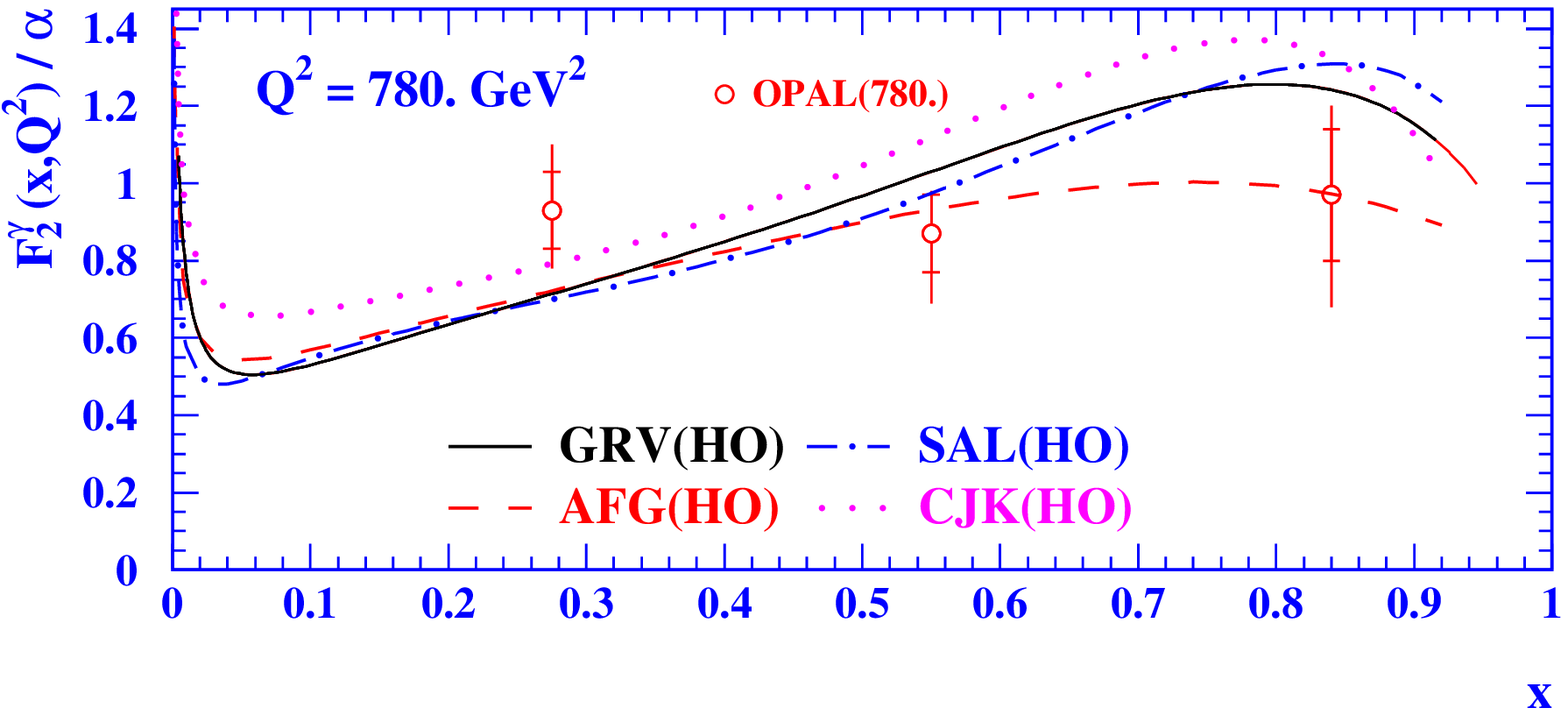}}
 \caption{Parametrisations of \ftxq compared to LEP data at 
          $\qsq=67.2$ and $780$ \gevsq.}
 \label{Fig:fig06}
\end{figure}
%
 The CJK parametrisation is based on all available \ftxq data for
 $\qsq>1$~\gevsq including the TPC/2$\gamma$ data and the preliminary
 DELPHI data taken at LEP1 energies. Various ways of treating the
 heavy quark, i.e.~$c,b$, contributions are explored by the CJK group,
 leading to various parametrisations.
 The parametrisation used in this review is the CJK NLO model, which
 is based on the ACOT($\chi$) variable-flavor number scheme. For
 brevity, it is denoted by CJK(HO).
 The parametrisation is evaluated in the DIS$_\gamma$ factorization
 scheme, the starting scale of the evolution as obtained from the fits
 is $Q_0^{2}=0.765$~\gevsq, and the strong coupling constant,
 $\alpha_{s}$, uses $\lamv=280$~MeV.  \par
 The AFG(HO) parametrisation is based on a subset of data, namely LEP1
 data at medium \qsq, including the preliminary DELPHI data. The heavy
 quarks are taken to be massless, however, $m_q^2/\qsq$ corrections to
 the direct component of \ftxq are included in the calculation.
 The AFG(HO) parametrisation is evaluated in the \msb factorization
 scheme, the starting scale for the evolution is $Q_0^{2}=0.7$~\gevsq,
 again as obtained from the fit, and $\lamv = 300$~MeV is used.
 \par
 Finally, the SAL(HO) parametrisation is based on a completely
 different theoretical concept, namely the assumption of the Gribov
 factorization, which relates the total $\gamma\gamma$ cross-section
 to the total $\gamma p$ and $pp$ cross-sections.
 At small values of $x$ the following relation between the proton and
 photon structure functions is obtained:
 $\ft=\frac{\sigma_{\gamma\,p}(W)}{\sigma_{p\,p}(W)}
 \cdot F_2^p\approx 0.43\cdot F_2^p$, 
 where the numerical value stems from the Donnachie-Landshoff
 parametrisation of the total cross-sections at large values of $W$.
 Consequently, the input data of \ftxq used, i.e.~all published \ftxq
 data except the TPC/2$\gamma$ data, are augmented by the ZEUS $F_2^p$
 results at $x<0.01$ and $\qsq<100$~\gevsq. 
 In the few overlapping regions the $F_2^p$ data are much more precise
 than the corresponding \ftxq data. In addition the $F_2^p$ results
 extend to much lower values of $x$.
 Consequently, the $F_2^p$ data determine the low-$x$ behavior of
 \ftxq.
 In addition, in an attempt to better constrain the gluon distribution
 of the photon, also ZEUS di-jet data measured in photo-production
 events are used.
 However, it turns out that in the present kinematical region of the
 data the sensitivity to the gluon from the photon is rather
 limited. The data are strongly dominated by contributions of quarks
 from the photon, while the fraction of events originating from gluons
 from the photon is very small.
 The relative division of data used for the fit for
 \ftxq/$F_2^p$/di-jet is about 7/5/1.
 For the treatment of heavy quarks the SAL group derives an
 interpolation between the fixed flavor number scheme at low values of
 \qsq and the zero-mass variable flavor number scheme at high values
 of \qsq.
 The SAL(HO) parametrisation is evaluated in the DIS$_\gamma$
 factorization scheme, the starting scale of the evolution is chosen
 to be $Q_0^{2}=2.0$~\gevsq, and $\alpha_{s}$ uses $\lamv = 330$~MeV.
 \par
 Despite their rather different theoretical framework all groups face
 a common difficulty, they have problems fitting the preliminary
 DELPHI data taken at LEP1 and/or LEP2 energies.
 Finally, this results in an inflation of the experimental error, or
 even in exclusion of this data. 
 \par
 The fact that all parametrisations are based on different theoretical
 prejudice and use different experimental input to their fits makes it
 even more interesting to compare their behavior to the experimental
 data.
 This comparison can be seen in Figures~\ref{Fig:fig05}
 and~\ref{Fig:fig06} for a number of \qsq values spanning almost the
 entire experimental range of the LEP data.
 The data shown at $\qsq=3.7 / 25.5 / 67.2 / 780$~\gevsq were partly
 used, (+), in the respective fit, and partly not, (-), where the
 corresponding patterns are for CJK(HO):~(+/-/+/+), for
 AFG(HO):~(+/-/-/-) and finally for SAL(HO):~(+/-/+/+).
%
\begin{figure}[thb]
 \centerline{\includegraphics[width=0.78\textwidth]{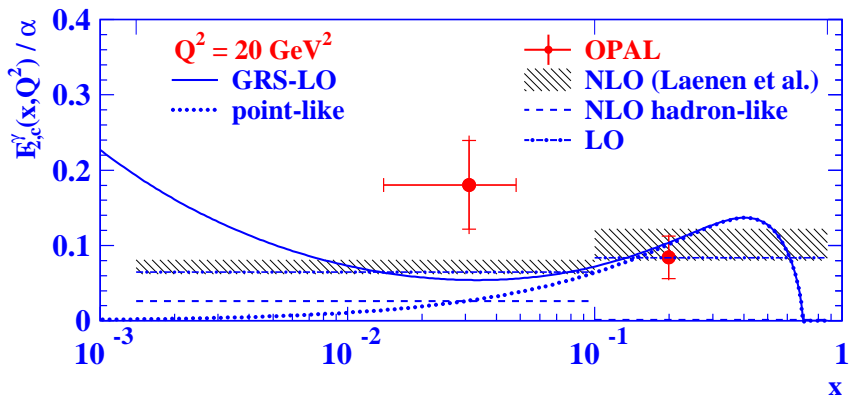}}
 \caption{The measurement of \ftc from OPAL.}
 \label{Fig:fig7}
\end{figure}
%
 Amongst the three new parametrisations CJK(HO) exhibits the steepest
 slope at low values of $x$, with increasing differences to the other
 parametrisations for increasing \qsq values.
 At medium values of $x$ the parametrisations are closer to each
 other, and at high values the differences increase again, with the
 AFG(HO) \msb parametrisation always yielding the lowest prediction.
 For comparison, the old GRV prediction generally lies in the middle
 of the new predictions, but for the lowest $x$ values at
 $\qsq=3.7$~\gevsq.
 Overall there is good agreement of the new parametrisations with the
 data at $\qsq=3.7 / 67.2 / 780$~\gevsq given the experimental
 uncertainties.
 The largest differences are seen for the data at $\qsq=25.5$~\gevsq,
 which have not been used by any of the fits and which have rather
 small uncertainties assigned.
 Here the most notable difference to the data is at low values of $x$
 when compared to the CJK(HO) prediction, which is significantly
 higher than the data. 
 \par
 Not only the inclusive structure function \ftxq has been obtained
 experimentally, but also its charm component, \ftc, has been
 measured~\cite{OPALPR354}.
 The charm part has been identified from the inclusive data by
 selecting charmed D mesons.
 The analysis makes use of the small phase space of the pion in the
 decay $D^\star \to D^\circ \pi$, which leads to a peaking structure in
 the distribution of the mass difference of the $D^\star$ and $D^\circ$
 mesons.
 The result for \ftc in two bins of $x$ and unfolded to
 $\qsq=20$~\gevsq is shown in Figure~\ref{Fig:fig7} in comparison to
 several theoretical predictions.
 Shown are the purely perturbative calculations
 from~\cite{LAE-9401,LAE-9602} at leading and next-to-leading order,
 NLO, and for the two data bins of $x$.
 This clearly shows that NLO corrections to \ftc are small.
 In addition shown is the functional form of the leading order
 GRS~\cite{GLU-9501} parametrisation for both the full \ftc and the
 point-like part alone.
 The separation in $x$ of the data has been such as to experimentally
 separate the point-like part, concentrated at large values of $x$,
 from the hadron-like part, dominating at low values of $x$, as can be
 seen e.g.~by comparing to the GRS curves.
 Figure~\ref{Fig:fig7} demonstrates that the high $x$ region is
 adequately described by the point-like NLO prediction with only
 $\alpha_{s}$ and the mass of the charm-quark, $m_{\mathrm{c}}$, as
 free parameters.
 The behavior at low values of $x$ is experimentally less well
 constrained.
 However, it can not be accommodated by the point-like part alone,
 e.g.~as given by the GRS parametrisation, thereby suggesting a
 non-vanishing hadron-like part at low values of $x$ also for \ftc.
 The uncertainty on the measurement at low values of $x$ is relatively
 big, but largely dominated by statistical uncertainties (inner error
 bars), so a measurement of \ftc by the other LEP experiments is
 highly desirable. 
 \par
%
\begin{figure}[thb]
 \centerline{\includegraphics[width=0.50\textwidth]{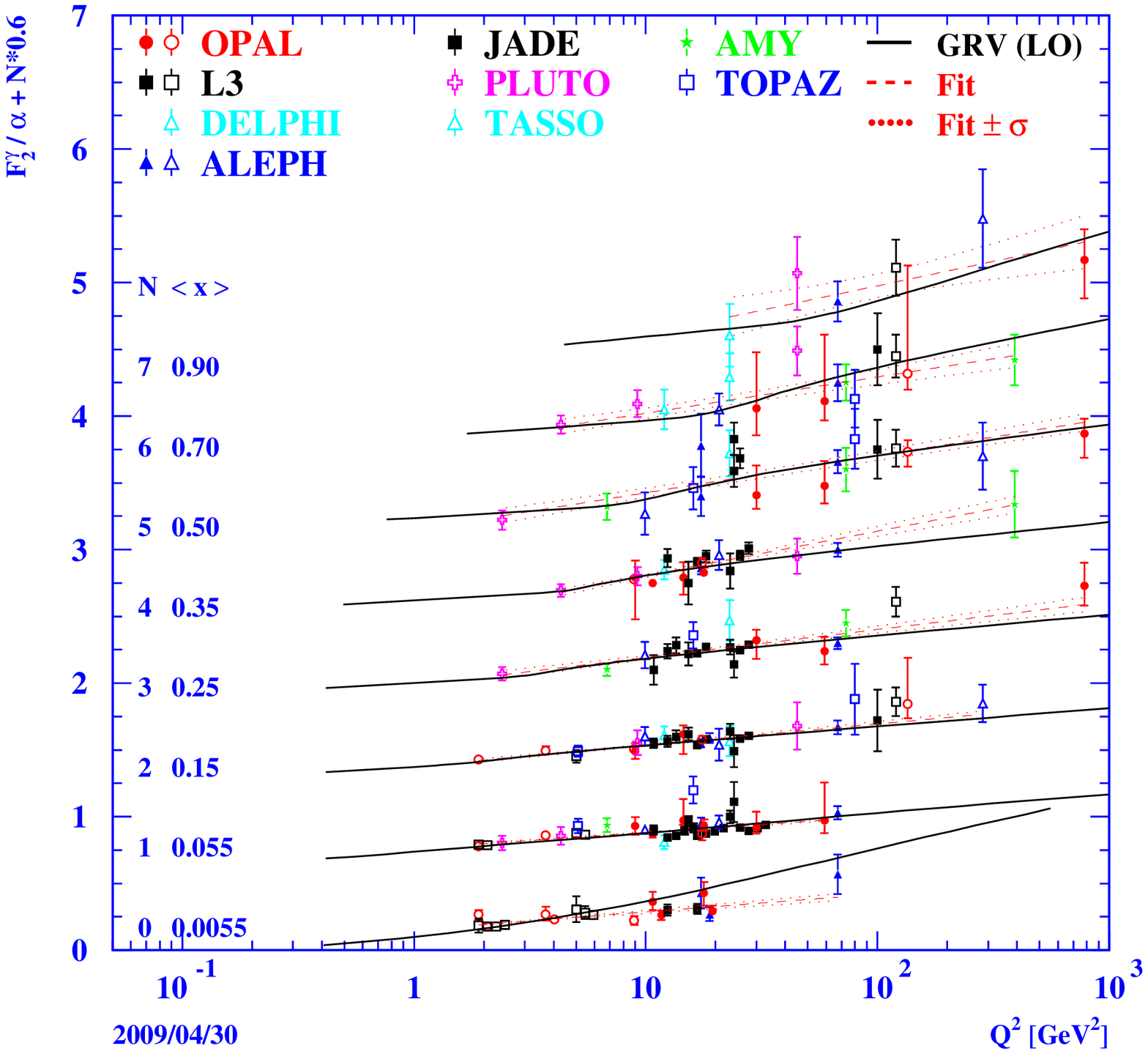}
             \includegraphics[width=0.47\textwidth]{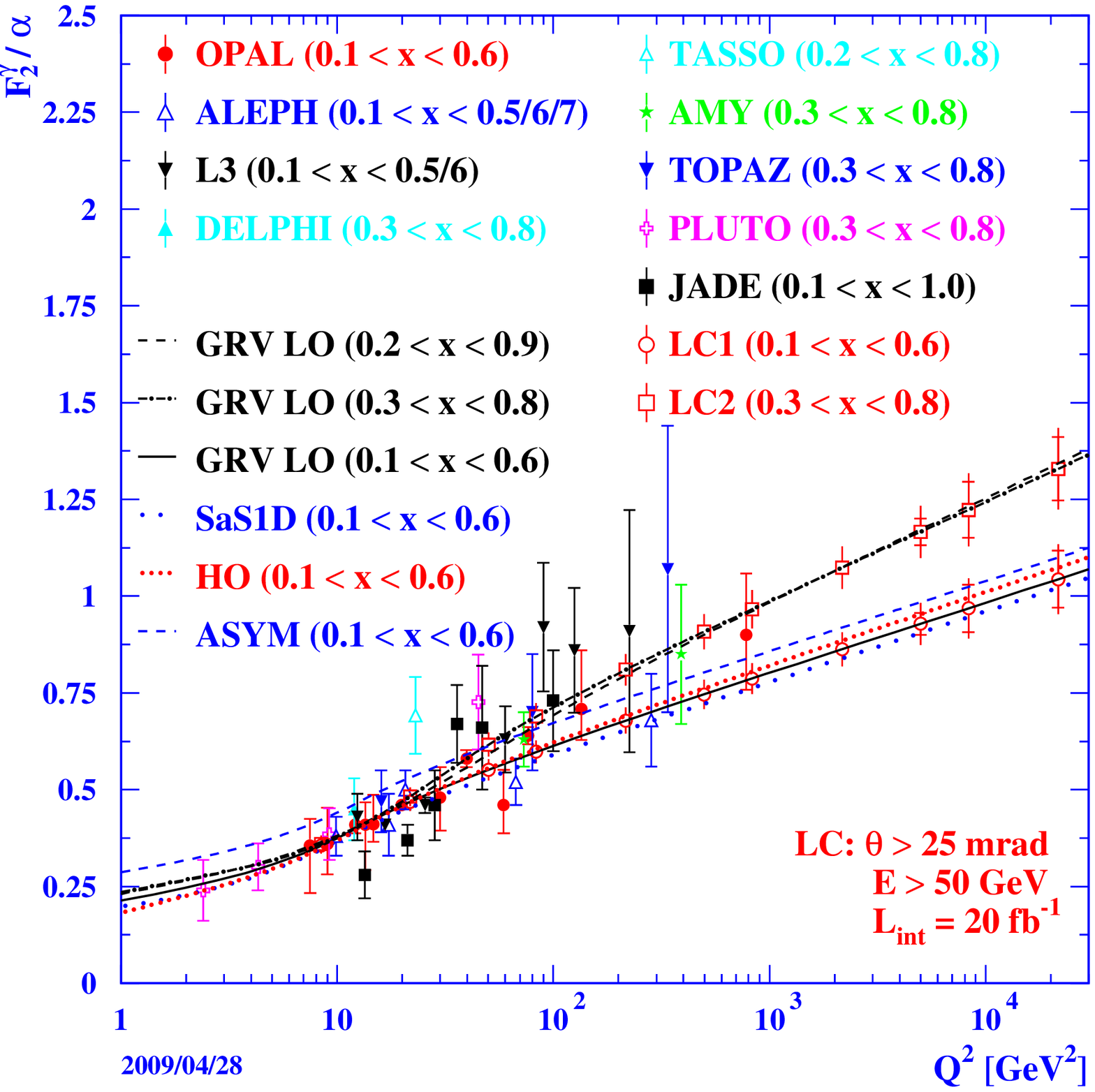}}
 \caption{Positive scaling violations of \ftxq for various regions in $x$ (left),
          and prospects for a measurement of the \qsq evolution of \ftxq at 
          an ILC (right).}
 \label{Fig:fig08}
\end{figure}
%
 One key feature of \ftxq is the logarithmic behavior with \qsq as
 predicted by perturbative QCD.
 It is the point-like contribution discussed above that results in
 positive scaling violations of \ftxq for all values of $x$, in
 contrast to the proton, which exhibits negative scaling violations at
 high values of $x$, due to gluon radiation, and positive scaling
 violations at low values of $x$, due to pair creation of
 quark--anti-quark pairs.
 See~\cite{NIS-9904} for a detailed assessment of this issue.
 \par
 The positive scaling violations of \ftxq for all values of $x$ is
 born out by the data as can be seen from
 Figure~\ref{Fig:fig08}(left). 
 The data are displayed as a function of \qsq in bins of $x$, where
 each data point is shown at its nearest average $x$ value chosen from
 the list on the left. In addition, for better visibility, the data
 points for the various bins in $x$ are separated by constant offsets,
 N.
 Linear fits to the data of the form $\ftq = a + b \ln \qsq$ have been
 performed. The fitted values of $b$ are significantly above zero for
 all bins of $x$, and a clear trend for increasing slope with
 increasing values of $x$ is observed.
 \par
 What about the future of \ftxq after LEP. There are two obvious
 candidates for future measurements, a short term opportunity is the
 measurement at the B-factories, where the Babar and Belle experiments
 are operating. The longer term option is the measurement of \ftxq at
 an International Linear Collider, ILC.
 The general prospects for Two-Photon physics at an ILC can be found
 in~\cite{NIS-9802}. 
 The higher beam energy and luminosity available at the ILC compared
 to LEP will allow to extend the available phase in \qsq by about two
 orders in magnitude.
 For a detailed investigation of neutral current interactions
 see~\cite{VOG-9901}.
 As an example, the measurement of the \qsq evolution of \ft at medium
 values of $x$ at an $\mathrm{e}^+\mathrm{e}^-$ collider is shown in
 Figure~\ref{Fig:fig08}(right).
 At the ILC also novel features can be investigated like the
 measurement of the flavor decomposition of \ft by exploring the
 exchange of $W$ and $Z$ bosons~\cite{GEH-9901}.
%
%
\subsection{Summary}
\label{sec:summa}
 The measurement of photon structure functions is an interesting field
 of research. Unfortunately, experimentally it has come to a halt
 after the shut-down of LEP, since so far it has not been pursued at
 the B-factories and the prospects for an ILC are still uncertain.
 \par
 Up to now, a wealth of data has been analyzed both in terms of the
 QED structure, and for the hadronic structure of the photon.
 In this short review only a part of the investigations could be
 discussed in detail.
 Concerning the QED structure, \ftqed was investigated, as well as
 additional structure functions from azimuthal correlations and the
 interactions of two virtual photons.
 For the hadronic structure the emphasis is on \ftxq and especially
 its behavior at low values of $x$ and the logarithmic scaling
 violations with \qsq.
 In addition, the charm contribution \ftc has been measured, and the
 interaction of two virtual photons were investigated.
 \par
 I strongly hope that the future will bring us additional information
 from the B-factories and an ILC.
%
%
\section*{Acknowledgments}
 I like to thank the organizers for this interesting conference, and
 for giving me the opportunity to refresh my memory on this interesting
 field of research.
%
\begin{footnotesize}

\end{footnotesize}
%
\end{document}